\documentclass[twocolumn,showpacs,amsmath,amssymb,prl]{revtex4}

\usepackage{graphicx,epsf}
\usepackage{bm}

\begin{document}

\title{Giant oscillations of energy levels in mesoscopic
superconductors}

\author{N. B. Kopnin $^{(1,2)}$}
\author{A. S. Mel'nikov $^{(3,4)}$}
\author{V. I. Pozdnyakova $^{(3)}$}
\author{D. A. Ryzhov $^{(3)}$}
\author{I. A. Shereshevskii $^{(3)}$}
\author{V. M. Vinokur $^{(4)}$}
\affiliation{$^{(1)}$ Low Temperature Laboratory, Helsinki
University of
Technology, P.O. Box 2200, FIN-02015 HUT, Finland,\\
$^{(2)}$ L. D. Landau Institute for Theoretical Physics, 117940
Moscow, Russia\\
 $^{(3)}$ Institute for Physics of Microstructures,
Russian Academy of Sciences, 603950, Nizhny Novgorod, GSP-105,
 Russia,\\
$^{(4)}$ Argonne National Laboratory, Argonne, Illinois 60439 }

\date{\today}

\begin{abstract}
The interplay of geometrical and Andreev quantization in
mesoscopic superconductors leads to giant mesoscopic oscillations
of energy levels as functions of the Fermi momentum and/or sample
size. Quantization rules are formulated for closed quasiparticle
trajectories in the presence of normal scattering at the sample
boundaries. Two generic examples of mesoscopic systems are
studied: (i) one dimensional Andreev states in a quantum box, (ii)
a single vortex in a mesoscopic cylinder.
\end{abstract}
\maketitle

\begin{figure}[t]
\centerline{\includegraphics[width=0.65\linewidth]{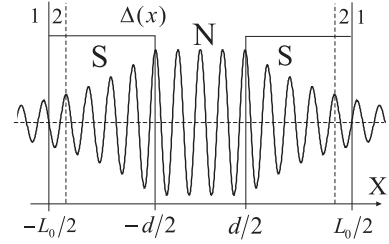}} %
\caption{Andreev states in an SNS structure placed in a quantum
box of length $L_0$. Solid (1) and dashed (2) lines show the
positions of the box boundaries for resonance and off-resonance
situations, respectively.}\label{fig-box}
\end{figure}

A normal cavity in a superconductor sample confines normal
carriers due to their Andreev reflection from the walls formed by
the superconductor order parameter. In the present Letter we show
that Andreev levels in samples with sizes comparable to the
coherence length exhibit giant mesoscopic oscillations as
functions of the Fermi momentum $k_F$ and/or sample dimensions
with an amplitude that substantially exceeds the interlevel
spacing they would have in bulk samples. For illustration, let us
compare the effects of geometrical confinement for bound states in
normal and superconducting systems. Consider  one dimensional
motion of a particle in a potential well of finite depth. The
particle wave function oscillates as $e^{ik x}$ in classically
accessible- and decays exponentially in forbidden regions
respectively. Placing the entire system into a quantum box of the
size $L_0$ larger than the width of the well, $d$, makes the wave
function to vanish at the box boundaries. This is equivalent to a
decrease in the effective width of the well and results in a
slight modification of the bound states. Let us now take a
sandwich-like structure of a total thickness $L_0$ where the
normal (N) slab of a thickness $d$ is confined between two
superconducting (S) layers with a certain order parameter phase
difference $\phi$ between them (see Fig.~\ref{fig-box}), and
consider the effect of the geometrical confinement on Andreev
states with energies $\epsilon$ below the superconducting gap
$\Delta$. As before, the particle/hole wave functions in the N
region oscillate with $k =k_F\pm \epsilon /\hbar v_F$, where $v_F$
is the Fermi velocity. However, in contract to the previous
example, the wave functions oscillate further into the S layers
with the period of $2\pi/k_F$ and with the amplitude slowly
decaying on the scale of the superconductor coherence length
$\xi_0 = \hbar v_F/\Delta$. Only when the particle in the box is
exactly in the geometrical resonance, $k_FL_0=\pi n$, i.e., when
one of its normal-state energy levels coincides with the Fermi
level, the Andreev states do not feel the external boundaries, see
Fig.~\ref{fig-box}. When the particle is out of resonance, $\sin
(k_FL_0)\sim 1$, to satisfy the zero boundary conditions an
adjustment of either the wave vector in the N region by $\delta k
\sim \delta \epsilon /\hbar v_F\sim 1/d$ (for $d>\xi_0$) or of the
phase shift between the electrons and holes by $\delta
\arccos(\epsilon/\Delta)\sim 1$ (for $d\lesssim \xi_0$) is needed.
Thus, the deviations in energy are large and may compare to
$\Delta$ for $d\sim \xi_0$.

We thus see that the interplay between geometrical and Andreev
quantization results in  giant oscillations of the Andreev levels
as functions of $k_F L_0$ with an amplitude of the order of
$\Delta$. The amplitude decreases exponentially as a function of
the distance from the external boundary to the Andreev turning
point of a particle at the NS interface. This oscillatory
phenomenon can be viewed as a generalization of geometrical
effects caused by the presence of impurity atoms in vortex cores
\cite{LarkOvch98} and Tomasch oscillations in films and
finite-size type-II superconductors \cite{Tomasch,Janko} but with
the enormously amplified magnitude. The mesoscopic oscillations
strongly affect both thermodynamic and transport properties of
nanoscale superconductors that are the focus of current
experimental and theoretical research (see
\cite{contact,Janko,multi} and references therein). Interference
effects of similar origin have been previously studied for hybrid
NS systems with multiple semi-transparent insulator potential
barriers \cite{Galaktionov} (see also \cite{contact} for a
review).

{\it Closed trajectories}.-- The one dimensional physics in a
quantum box is related  to the concept of closed trajectories
whose impact on the validity of the quasiclassical description of
superconductivity is discussed in \cite{shelankov2001}. Making use
of this concept enables one to generalize the above picture to
higher dimensions. A standard semiclassical approach to
superconductors is formulated for the quasiparticle motion along
the beams ${\bm \nabla} S$ where $S$ is the normal-state eikonal,
$|{\bm \nabla}S|=k_F$. The wave function of fermionic excitations
has two components in the particle-hole space $\hat \Psi =\left( u
, v\right)$ and can be written as $ \hat \Psi =\hat \psi e^{iS} $,
where $\hat \psi$ is the envelope function varying slowly over the
Fermi wavelength. Generally, the phase $S$ gained along open
trajectories does not affect excitation characteristics such as
energy levels or density of states (DOS). However, the closed
trajectories formed due to boundaries, impurity potentials, and/or
in an applied magnetic field, do influence the energy spectrum.

Consider a closed trajectory of the length $L$. The wave function
$\hat \Psi$ should be single valued, which gives
\begin{equation}
\hat \psi (s) = \hat \psi (s+L) e^{iS (L)} \ ,
\label{s-valued}
\end{equation}
where $S(L)=\oint {\bm \nabla}S\cdot d{\bf r}= k_F L$ is
independent of the arc length $s$ along the trajectory. Equation
(\ref{s-valued}) suggests that the initial problem is equivalent
to the problem of an unbounded motion of a particle in the
periodic gap potential $\Delta (s)=\Delta(s+L)$ for the proper
choice of the quasimomentum $q$ (see below). In the latter case
the wave function satisfies the Bloch theorem
\begin{equation}
\hat \psi _q(s+L) = e^{iqL}\hat \psi_q (s)  \ , \label{q-momentum}
\end{equation}
while the energy is a periodic function of $q$: $\epsilon
(q+2\pi/L)=\epsilon (q)$. Comparing Eqs.~(\ref{s-valued}) and
(\ref{q-momentum}) we find $q=[2\pi M-S(L)]/L$, where $M$ is a
large integer chosen such that $q$ belongs to the first Brillouin
zone. Consider two examples: (i) States with $\epsilon > \Delta
=const$ will have the spectrum $\epsilon^2 = \Delta^2 + \hbar^2
v_F^2 (2\pi M^\prime /L - k_F)^2$ where $M^\prime =M-N$, $N$ being
the number of the energy band. This spectrum results in Tomasch
oscillations \cite{Tomasch,Janko}. (ii) For sub-gap states the
correspondence between the spectrum in a mesoscopic superconductor
and in a bulk sample can be easily established for $L\gg\xi_0$.
Let $\epsilon^{(0)}$ be an Andreev bound state for $L\rightarrow
\infty$. The tight binding approximation in the equivalent
periodic problem yields
\begin{equation}
\epsilon=\epsilon^{(0)}+\delta ( \cos\alpha +C) \ ,
\label{energy-tight}
\end{equation}
where the band width $\delta \sim \Delta e^{-\lambda L}$ is
proportional to the exponential overlap of the decaying functions
for the sub-gap states, $\lambda (\epsilon )\sim \xi_0 ^{-1}$, and
$\alpha =qL+\beta$. The parameters $\beta $ and $C$ depend on the
particular problem. Since $\cos \alpha =\cos (k_FL-\beta)$, the
energy level oscillates rapidly as a function of $k_F L$. The
amplitude of oscillations can well exceed the value of
$\epsilon^{(0)}$ itself, provided the loop length $L$ is not much
larger than $\xi _0$.

The relative contribution of such oscillations to bulk properties
depends on the relative weight of closed trajectories allowed by
the particular sample geometry. Below we focus on two problems
where these contributions are critical: (i) one dimensional (1D)
Andreev states in a quantum box (this problem has been discussed
briefly in the introduction), and (ii) energy states in a vortex
core placed in a clean mesoscopic cylinder of a finite radius. In
both cases the dimensions of the system are assumed comparable to
$\xi_0$.

{\it 1D Andreev bound states in a quantum box}. -- Consider a
quantum point contact which is transparent to a few modes $N_c$
passing from one superconducting lead to another. We further
assume that these modes are localized within the device by
specular reflections at the boundaries of the leads which are
separated by a distance $L_0$. The leads contain also large
numbers of other modes $N_{lead}\gg N_{c}$, which take part in the
superconducting pairing. The leads can be connected, via some of
the modes $N_{lead}$, to an external superconducting circuit to
control the phase difference $\phi$ between them. A possible
realization of this device is an adiabatic constriction of the
type discussed in \cite{been1}. For the modes $N_{c}$ that pass
through the constriction but are confined within the box of the
size $L_0$, the quantum mechanical problem corresponds exactly to
that in Fig.~\ref{fig-box} for the limit $d\ll \xi _0$.

We assume a step-like gap potential $\Delta (x)=\Delta_0 e^{i \,
{\rm sign}(x)\, \phi /2}$. The confinement couples the states with
opposite momenta and creates a closed trajectory loop of the
length $L=2L_0$. In the S region one has the waves $e^{i\tilde
q_\pm x}$ and $e^{-i\tilde q_\pm x}$, where $\tilde q_\pm = k_x\pm
i \lambda $, $ \lambda =\sqrt{\Delta_0^2 -\epsilon ^2}/\hbar v_x$,
$k_x$ and $v_x$ are the particle momentum and velocity projections
on the $x$ axis. We choose here $|\epsilon|<\Delta_0$, however,
the same expressions hold also for $|\epsilon|>\Delta_0$ with an
imaginary $\lambda$. Matching these wave functions yields the
dispersion relation
\begin{equation}
\epsilon^2 =\Delta_0^2\left[1-{\cal T} \sin ^2(\phi /2)\right] \ .
\label{energy-phi}
\end{equation}
The transmission coefficient ${\cal T} =(1+A)^{-1}$, and $ A=
\sin^2 (k_xL_0)/\sinh^{2} (\lambda L_0) $. For $L_0\gg \xi _0$ and
$\epsilon <\Delta_0$, Eq.~(\ref{energy-phi}) looks like
Eq.~(\ref{energy-tight}) in accordance with the general arguments
above. Equation (\ref{energy-phi}) has a familiar form
\cite{contact} of the spectrum for a contact with the double
barrier of strength $A$. It describes mesoscopic fluctuations and
accounts for the resonance transmission at $\sin (k_xL_0)=0$.
However, the effect of spectrum modifications is not restricted to
mere renormalization of the barrier strength. Qualitatively new
features appear due to the energy dependence of the transmission
coefficient ${\cal T}$. Because of the geometrical quantization,
the spectrum for $|\epsilon |>\Delta_0$ is no longer a continuum
and cannot be separated from the phase-dependent sub-gap state;
instead, we obtain a set of $\phi$-dependent discrete levels in
the entire energy range. For a short box $L_0\lesssim \xi _0$, the
lowest energy level is
\[
\epsilon_{0}^2(\phi) = \Delta_0^2 [\cos^2(\phi /2)+ \left(\hbar
v_x/\Delta_0 L_0\right)^2 \sin^2 (k_xL_0)] \ .
\]
For $|\epsilon |\gg\Delta_0$, the levels transform into the
$\phi$-independent spectrum in a normal-metal box: $(\hbar
k_x\pm\epsilon_n/v_x)L_0=\pi n \hbar$, where $n$ is an integer.
Each $\phi$-dependent level provides an oscillatory contribution
to the supercurrent \cite{been}:
\begin{equation}
 I_n(\phi) = -\frac{2e}{\hbar}
\frac{d \epsilon_{n}(\phi)}{d \phi}
\tanh\frac{\epsilon_{n}(\phi)}{2T} \ , \label{supercurrent}
\end{equation}
where only $\epsilon_n >0$ are taken. Note that the supercurrent
in Eq.~(\ref{supercurrent}) is transported by the modes which, in
the normal state, are localized and do not carry current. In the
superconducting state, however, the current along the localized
modes $N_{c}$ appears due to the conversion over distances
$\sim{\rm min} \{L_0, \xi_0\}$ of the supercurrent from
delocalized modes $N_{lead}$. The analysis of
Eq.~(\ref{energy-phi}) shows that the strongest $\phi$ dependence
is realized for the lowest energy state which thus dominates the
current.

{\it Vortex core states in a mesoscopic superconductor}. -- We
consider now low energy core states in a single vortex introduced
into the center of a mesoscopic cylinder  of a radius $R\gtrsim
\xi _0$ with the quasiparticle mean free path $\ell \gg R$.
Bogoliubov-de~Gennes (BdG) equations read
\[
\left(-\frac{\hbar ^2}{2m}{\bm \nabla}^2 -E_F +\hat
\Delta\right)\hat \Psi = \epsilon \hat\sigma _z \hat \Psi \ ,
\]
where $\hat \Delta =|\Delta (r)| e^{i\hat \sigma _z\phi} i\hat
\sigma _y$, $\hat \sigma _i$ are Pauli matrices, and $r,\phi, z$
are cylindrical coordinates with the $z$ axis parallel to the
cylinder axis. The gap $|\Delta (r)|$ saturates at $\Delta _0$ far
from the vortex axis. The vector potential here is negligible for
an extreme type-II superconductor. We look for a solution $\hat
\Psi =e^{i\hat\sigma_z\phi /2 +i\mu \phi }\hat{\cal U}$ with a
given half-integer angular momentum $\mu$. If the superconducting
cylinder is surrounded by an insulator, the boundary condition
requires $ \hat{\cal U}(R,z)=0 $.

\begin{figure}[t]
\centerline{\includegraphics[width=1.0\linewidth]{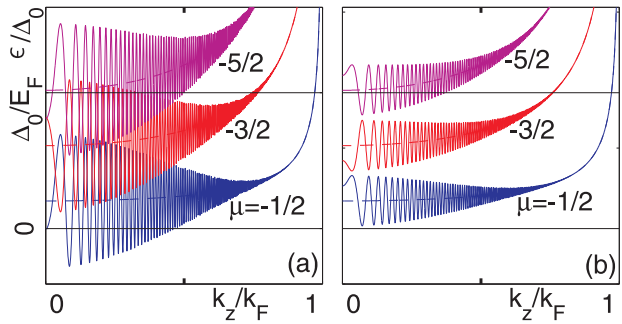}} %
 \caption{
 \label{fig:spectrum}
The energy spectra (solid lines) for a vortex in a mesoscopic
cylinder with (a) $R/\xi_0=3.5$ and (b) $R/\xi_0=4.0$  as
functions of $k_z$. The CdGM energy spectra are shown by the
corresponding dashed lines.}
\end{figure}

We find the energy spectrum both analytically and numerically. For
numerical computations we use a matrix representation of the BdG
operator in the basis of the eigenstates with a given momentum
$k_z$ along the $z$ axis for a normal-metal cylinder of the radius
$R$. We truncate the infinite matrix keeping the number of
eigenstates larger than the number of propagating modes in the
normal-metal waveguide. The obtained $k_z$-dependent matrix is
diagonalized yielding the energy spectrum for a vortex. We
approximate the gap by $|\Delta(r)|=\Delta_0
r/\sqrt{r^2+\xi_v^2}$, choosing $\xi_v=\xi_{0}$ without the loss
of generality. The calculated spectra do not depend qualitatively
on the exact shape of $|\Delta(r)|$. Shown in
Fig.~\ref{fig:spectrum} are typical energy spectra calculated for
a realistic material parameter $\Delta_0/E_F =0.01$.

The analytical description is based on a standard quasiclassical
scheme \cite{CdGM} modified to take account of the proper phases
\cite{Skvortsov} of radial waves for small values of $\mu$:
\[
\hat{\cal U}= e^{ik_z z}H^{(1)}_{\mu+\hat\sigma_z/2}
\left(k_r r\right)\hat w^{(+)}
+e^{ik_z z}H^{(2)}_{\mu+\hat\sigma_z/2}\left(k_r r\right)\hat w^{(-)} \ ,
\]
where $H_l^{(1,2)}$ are the Hankel functions, $k_r^2+k_z^2=k_F^2$,
and $\hat w=\left(w_{1},\, w_{2}\right)$ are slow functions of
$r$. Solving equations for the envelopes $\hat w^{(\pm )}$ one can
construct functions $\hat w^{>}$ and $\hat w^{<}$ decaying at
different ends of the trajectory passing by the vortex
\cite{KMV03}. The wave function in a cylinder is the superposition
$ \hat w =A_>\hat w^{>}+ A_<\hat w^{<} $. The boundary condition
at $r=R$ couples the incoming and outgoing waves which leads to
formation of a closed trajectory loop for one-dimensional motion
along $r$. The solvability condition of the two linear homogeneous
equations for two constants $A_<$ and $A_<$ gives the bound state
energy $\epsilon (q)\equiv \epsilon _\mu (k_r)$ in the form of
Eq.~(\ref{energy-tight}) where $\delta =\Delta_0
/(\Lambda\cosh[2K(R)])$, $\alpha =2k_rR-\pi \mu +\pi /2$, $C=0$,
and $\epsilon ^{(0)} \equiv \epsilon _\mu ^{(0)} =-\omega (k_r)\mu
$ is the Caroli--de Gennes--Matricon (CdGM) energy \cite{CdGM} for
a vortex in a bulk superconductor,
\begin{equation*}
\omega (k_r) =\frac{2m\Delta _0}{\hbar ^{2} k_r^{2}\Lambda}
\int_0^{\infty}\left(|\Delta (r)|/r\right) e^{-2K(r )}\, dr \ ,
\end{equation*}
\begin{equation*}
K(r)=\frac{m}{\hbar ^{2} k_r}\int _0^r |\Delta (r^\prime )|\,
dr^\prime   , \; \Lambda =\frac{2m\Delta _0}{\hbar ^{2}k_r}\int
_0^\infty e^{-2K (r)}\, dr \ .
\end{equation*}

These analytical expressions are in a very good agreement with our
numerical results. The last term in Eq.~(\ref{energy-tight})
describes the mesoscopic level fluctuations. Their amplitude $
\delta (k_r)$ is much larger than the CdGM level spacing $\omega
(k_r)\sim \Delta _0^2/E_F$ if $R$ is not exceedingly larger than
$\xi_0$. Different levels can cross each other because they belong
to different angular momenta $\mu$. The fluctuating levels also
can cross zero for not very high $\mu$. The amplitude
$\delta(k_r)$ decreases as $k_z$ approaches $k_F$ since $\cosh[2
K(R)]$ in it grows exponentially as $k_r$ decreases. Note that in
a layered two-dimensional superconductor the amplitude
$\delta(k_F)$ is constant corresponding to $k_z=0$. For later use,
we define the critical radius $R_c$ for which the maximum
amplitude of oscillations $\delta(k_F)$ is equal to half of the
maximum distance between the CdGM states $\omega_0\equiv \omega
(k_F)$. According to Fig.~\ref{fig:spectrum}(b) the radius $R=4\xi
_0$ is slightly larger than $R_c$ for $\Delta_0/E_F =0.01$.

{\it Density of vortex core states.}-- Consider a sample radius
$R\lesssim R_c$ such that $\delta (k_F)\gg \omega_0$. According to
Fig.~\ref{fig:spectrum}(a), we divide the range $0<k_r<k_F$ into
the region of large oscillations $k^* _r<k_r<k_F$, where $\delta
(k_r) \gtrsim \omega (k_r)$, and the region $0 <k_r<k_r^*$, where
levels are smooth functions of $k_r$ close to the CdGM form
$\epsilon_\mu ^{(0)}(k_r)$. The total DOS is a sum $\nu =\nu
_1+\nu _2$ of contributions from the both regions. In the region
$k^* _r<k_r<k_F$, a large number of levels with different $\mu$
cross the constant energy line many times so that the continuous
approximation is appropriate. Therefore, $\nu _1$ is equal to the
CdGM DOS averaged over a large number of angular momentum
eigenstates:
\begin{equation}
\nu _1(k_r^*) =\frac{1}{\pi}\int _{k_r^* }^{k_F} \frac{k_r\,
dk_r}{\omega (k_r)\sqrt{k_F^2-k_r^2}} \  \label{DOS1}
\end{equation}
(per spin projection). It provides a background zero-energy DOS
due to mesoscopic level fluctuations. The term $\nu _2(\epsilon)$
is a sum of peaks positioned at $\epsilon=\epsilon^{(0)} _\mu
(k_r^*)$ with the energy period $\omega(k_r^*)$ larger than the
CdGM minigap $\omega _0$. In the limit $\epsilon\gg \omega(k_r^*)$
the term $\nu_2(\epsilon)$ saturates at the value determined by
the same integral as in Eq.~(\ref{DOS1}) but taken within the
limits $0$ to $k_r^*$. Therefore, the total DOS saturates at the
averaged CdGM DOS $\nu_0\equiv \nu _1(0)$.

\begin{figure}[t]
\centerline{\includegraphics[width=1.0\linewidth]{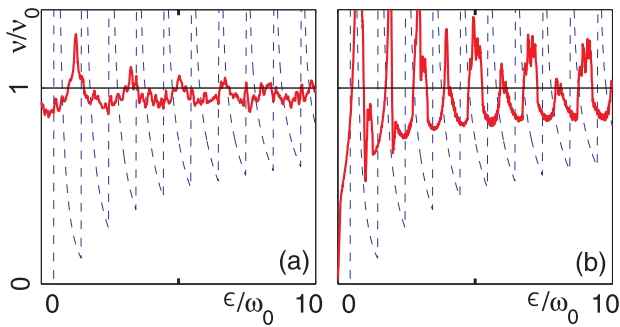}}
\caption{Solid lines: The DOS for (a) $R/\xi_0=2.5$ and (b)
$R/\xi_0=4$. The CdGM DOS is shown by dashed line.}\label{fig:DOS}
\end{figure}

If $R>R_c$ but the amplitude of oscillations is still comparable
to $\omega _0$, there are many roots of $\epsilon _\mu
(k_r)=\epsilon$ for a given $\mu$, such that the effective minigap
is $\epsilon _{\rm min}= \omega _0/2-\delta (k_F)$. It vanishes
for $R=R_c$.

To simplify our equations we approximate the CdGM interlevel spacing
 as $ \omega(k_r)\approx (k_F/k_r)\omega _0 $ while
$K(R)\approx (R/\xi _0)(k_F/k_r)$ with $\Lambda \approx 1$. We
find $ k_r^*/k_F =R/R_c $ and $R_c =(\xi_0/2)\ln \left(\Delta _0
/\omega _0 \right)$. The period of DOS oscillations is thus $
\omega(k_r^*) =\omega _0 R_c/R$. The background DOS is
\[
\nu_1=\nu _0 \left[1-(2/\pi)\left(\arcsin \rho -\rho \sqrt{1-\rho
^2} \right)\right]\ ,
\]
where $\nu _0=k_F/4\omega _0$ and $\rho =R/R_c$. It vanishes for
$\rho =1$.

We calculated the DOS numerically using the obtained analytical
expressions for the energy spectrum. To exclude a large number of
van Hove singularities the DOS was averaged over a small energy
interval $\delta\epsilon=0.1\omega_0$. The results for $R<R_c$ and
$R\approx R_c$ shown in Fig.~\ref{fig:DOS}(a) and (b),
respectively, are in good agreement with the above analytical
estimates. In particular, the period of oscillations in
Fig.~\ref{fig:DOS}(a) is approximately 1.5 times larger than the
period $\omega _0$ for the CdGM DOS in the bulk, which agrees with
the value of the cylinder radius $R\approx R_c/1.5$. At the same
time, the period in Fig.~\ref{fig:DOS}(b) almost coincides with
$\omega _0$; in addition, the minigap here vanishes. These
features well correspond to $R$ being close to $R_c$.

To summarize, we predict a profound effect of geometrical
quantization on Andreev states in mesoscopic superconductors,
which exhibit giant oscillations as functions of the particle
momentum and the sample size. In particular, the geometrical
quantization results in appearance of zero energy modes for vortex
core states. We discussed the case of ideal sample surfaces,
however one expects all the essential conclusions to hold for
atomically smooth sample surfaces as well. The spectrum
oscillations can be observed by scanning tunnelling spectroscopy
with high energy resolution and by transport measurements in weak
links.



We thank A. Andronov, S. Sharov, and A. Shelankov for valuable
discussions. This work was supported, in part, by the US DOE
Office of Science, contract No. W-31-109-ENG-38, by Russian
Foundation for Basic Research, by Program ``Quantum Macrophysics''
of Russian Academy of Sciences, by Russian Presidential Program
under grant No. MD-141.2003.02, by Russian Science Support
Foundation, by the ``Dynasty'' Foundation, and by the Russian
Ministry of Science and Education.


\end{document}